\documentclass[12pt,oneside]{amsart}
\usepackage{geometry}                
\usepackage[parfill]{parskip}    
\usepackage{natbib} 					
\usepackage{booktabs}
\usepackage[colorlinks]{hyperref} 	
\hypersetup{  
citecolor=black, 						
linkcolor= blue} 						
\usepackage{graphicx}
\usepackage{amssymb}
\usepackage{amsfonts}
\usepackage{amsmath}

\begin{document}
\title{Monte Carlo-based tail exponent estimator.}
\date{}
  \author{Jozef Barunik* \hspace{10 mm} and \hspace{10 mm} Lukas Vacha** \\}
\thanks{* Corresponding author, Institute of Information Theory and Automation, Academy of Sciences of the Czech Republic, Institute of Economic Studies, Charles University, Prague, Email: barunik@utia.cas.cz}
\thanks{** Institute of Information Theory and Automation, Academy of Sciences of the Czech Republic, Institute of Economic Studies, Charles University, Prague, Email: vachal@utia.cas.cz} 
 \thanks{The authors would like to thank Martin Smid for many useful discussions and careful reading of the manuscript. 
Support from the Czech Science Foundation under Grants 402/09/0965, 402/09/H045 and 402/08/P207 and from Ministry of Education MSMT 0021620841 is gratefully acknowledged.}

\begin{abstract}
In this paper we propose a new approach to estimation of the tail exponent in financial stock markets. We begin the study with the finite sample behavior of the Hill estimator under $\alpha-$stable distributions. Using large Monte Carlo simulations we show that the Hill estimator overestimates the true tail exponent and can hardly be used on samples with small length. Utilizing our results, we introduce a Monte Carlo-based method of estimation for the tail exponent. Our proposed method is not sensitive to the choice of tail size and works well also on small data samples. The new estimator also gives unbiased results with symmetrical confidence intervals. Finally, we demonstrate the power of our estimator on the international world stock market indices. On the two separate periods of 2002--2005 and 2006--2009 we estimate the tail exponent. \\
\\
\textit{JEL: C1, C13, C15, G0} \\
\textit{Keywords: Hill estimator, $\alpha$-stable distributions, tail exponent estimation} 
\end{abstract}
\maketitle

\newpage
\section{Introduction}

Statistical analysis of financial data has been of vigorous interest in recent years, mainly among the physics community \citep{MantSt2000, BouchPot2001, Bauch2002, Manteg1999, Plerou2000, Stanley2000, Stanley2003}. One of the main reasons driving the research is the use of established statistical characteristics to better describe and understand real-world financial data.

Dynamics of financial markets is the outcome of large number of individual decisions based on heterogeneous information. Financial returns representing the interaction of the market participants have been assumed to be Normally distributed for a long time. The strongest argument supporting this assumption is based on the Central Limit Theorem, which states that the sum of a large number of independent, identically distributed variables from a finite-variance distribution will tend to be normally distributed. However, financial returns showed to have heavier tails, which is a possible source of infinite variance.

\cite{Mandel1963} and \cite{Fama1965} proposed stable distributions as an alternative to the Gaussian distribution model. Stable distributions were introduced by \cite{Levy1925}, who investigated the behavior of sums of independent random variables. Although we know other heavy-tailed alternative distributions (such as student's $t$, hyperbolic or normal inverse Gaussian), stable distributions are attractive for researchers as they are supported by the generalized Central Limit Theorem. The theorem states that stable laws are the only possible limit distributions for properly normalized and centered sums of independent, identically distributed random variables. A sum of two independent random variables having a L\'evy stable distribution with parameter $\alpha$ is again a L\'evy stable distribution with the same parameter $\alpha$. However, this invariance property does not hold for different values of $\alpha$. Observed stock market prices are argued to be the sum of many small terms, hence a stable model should be used to describe them. When $\alpha<2$, the variance of the stable process is infinite and the tails are asymptotically equivalent to a Pareto law, i.e., they exhibit power-law behavior. Stable distributions have been proposed as a model for many types of physical and economic systems as they can accommodate fat tails and asymmetry and fit the data well. Examples in finance and economics are given in \cite{Mandel1963}, \cite{Fama1965}, \cite{Embrecht1997} or \cite{Rachev2000}.

There have, however, been many applications of L\'evy stable distributions to empirical data sets which could raise doubts about the correctness of the tail estimate \citep{Lux1996, Voit2005,Podobnik2000}. There is a significant difference between the value of the estimated $\alpha$ (based on the whole data set) and the estimated tail exponent. The tail exponent is estimated only on an arbitrarily chosen part of the data \citep{Hill1975, Weron2001}. Since extreme observations of prices on financial markets are of great importance, this problem deserves further research. If $\alpha$ is underestimated, the occurrence of extreme events is overestimated. \cite{Weron2001} shows that the estimated tail exponent is very sensitive to changes in parameters and to the size of the data set, hence the estimates can be highly misleading. Simulations show that a large data set ($10^{6}$) is needed for identification of the true tail behavior. The logical step is to use high-resolution data analysis. \cite{Lux1996} was one of the first to use high-frequency data, doing so for analysis of the German stock market index. Several studies concerning estimation of stable distributions followed \citep{MantSt2000, Dacorogna2001, Voit2005}.

In our paper, we append an analysis of the finite sample properties of the Hill estimator to the discussions. Moreover, we introduce a Monte Carlo-based tail exponent estimation method. In the first part, we briefly discuss the basics of the Hill estimator as well as stable distributions and their tail behavior. In the second part, we provide the finite sample properties of the Hill estimator and discuss the implications for its use on real-world data. In the third part, we utilize the results and propose a tail exponent estimation method based on Monte Carlo simulations. Finally, we illustrate the power of our method of estimation on leading world stock market indices and conclude.

\section{Estimation of the tail exponent}
The simplest method of estimating the tail exponent $\alpha$ is log-log linear regression. This method is very sensitive to the sample size and the choice of the number of observations used in the regression. \cite{Weron2001} shows that it also yields biased estimates and cannot be used for financial applications. Popular method for estimating the tail exponent is also rank-frequency plot. \cite{GabaixIbragimov2009} proposed an extension of the rank-frequency plot with lower bias for small samples. Maximum likelihood method of estimation still belongs to most reliable techniques. For example \cite{Clauset2009} combine maximum-likelihood fitting with goodness-of-fit tests based on Kolmogorov-Smirnov statistic and likelihood ratios. Recently, many authors have studied the return intervals between consecutive price fluctuations above some volatility threshold. \cite{Podobnik2009} found a simple relationship between the mean return intervals $\bar{\tau_q}$ and the threshold $q$, $\bar{\tau_q}\propto q^{\alpha}$, where $\alpha$ is power-law exponent. As authors show, this approach can also be used for estimation of the power-law exponent for financial data.

Another widely used  tail exponent estimator, equivalent to maximum-likelihood fitting, based on order statistics is proposed in \cite{Hill1975}. \cite{Pickands75} and \cite{Dekkers89} show other variations of the Hill estimator. \cite{Mittnik98} provide a modification of the Pickands estimator using high-order approximation. A quantile method is used by \cite{McCulloch1996}. For more detailed discussion on tail estimators see \cite{Embrecht1997} and \cite{Resnick2007}.
 
The Hill estimator tends to overestimate the tail exponent of a stable distribution if the value of $\alpha$ is close to two and the sample size is not very large. For a more detailed discussion see \cite{Weron2001} and \cite{Embrecht1997}. Several researchers have used misleading estimators of the tail exponent $\alpha$ to conclude that various data sets had $\alpha>2$, i.e., the data is not stable. First of all, let us briefly introduce the Hill estimator.
\subsection{Hill estimator}

The most popular method for estimating the tail exponent $\alpha$ is the Hill estimator \citep{Hill1975}. The Hill estimator is used to estimate the tail exponent only, therefore it does not assume a parametric form for the entire distribution function.
 
Let's suppose $X_1, X_2,\ldots,X_n$ is a sequence of i.i.d. random variables with the distribution function $F(x)$. Furthermore, let's assume that $1-F(x)$ has a regularly varying upper tail with coefficient $-\alpha$.
\begin{equation}
P\left( X>x\right)=1-F(x)=x^{-\alpha }L(x),\hspace{1cm} x>0
\label{eq01}
\end{equation}
where the function $L(x)$ is slowly varying at infinity\footnote{If eq. \ref{eq01} is satisfied, the distribution function $F(x)$ belongs to the maximum domain of attraction of $\phi_\alpha$, $F\in MDA(\phi_\alpha)$, where $\phi_\alpha (x)=e^{-x^\alpha}$, $x>0, \alpha>0$. For a more detailed treatment of the extreme value theory see \cite{Embrecht1997} } (for more details see \cite{Resnick2007}). 

Equation \ref{eq01} indicates that the right tail of the distribution function $F(x)$ has the same asymptotic properties as the tail of the Pareto distribution, see \citep{WagnerMarsh2004}.

Let us define the order statistics $X_{(1)} \ge X_{(2)} \ge\ldots\ge X_{(n)}$. The Hill estimator of the tail exponent $\alpha$ is defined as:
\begin{equation}
\hat{\alpha}_H=\left(\frac{1}{k}\sum_{i=1}^{k}\log{X_{(n-i+1)}}-\log{X_{(n-k)}}\right)^{-1},
\end{equation}
where $k$ is a truncation (or smoothing) parameter which defines a subsample used for the estimation, $k<n$.

Many authors have studied the asymptotical properties of the Hill estimator. According to \cite{Mason1998} the Hill estimator is weakly consistent if 
\begin{equation}
k\rightarrow\infty, \hspace{1cm}\frac{k}{n}\rightarrow0 \mbox{  as  } n\rightarrow\infty.
\end{equation}

Furthermore, \cite{GoldieSmith87} proved the asymptotic normality of the Hill estimator, i.e.,
\begin{equation}
\sqrt{k}(\hat{\alpha}_H^{-1}-{\alpha}^{-1})\sim N(0,{\alpha}^{-2}).
\end{equation}
Consequently, $\hat{\alpha}_H$ is also approximately normal with mean $\alpha$ and variance $\alpha^{2}/k$. Moreover, \citep{Hsing1991} shows that the Hill estimator is asymptotically quite robust with respect to deviations from independence. For a more detailed discussion of the properties of the Hill estimator see \cite{Resnick2007}, \cite{Embrecht1997}.

\subsection{Choice of optimal $k$ in the Hill estimator}

In the classical approach, there is considerable difficulty in choosing the right value of the truncation parameter $k$, since it can influence the accuracy of the estimate significantly. The methods of choosing $k$ from the empirical data set are very often based on a trade-off between the bias and the variance of the Hill estimator. $k$ has to be sufficiently small to ensure the observations $X_{(1)} \ge X_{(2)} \ge\ldots\ge X_{(k)}$ still belong to the tail of the distribution. On the other hand, if $k$ is too small, the estimator lacks precision.

There are several methods for choosing the ``optimal'' value of $k$. The first possibility is to make a Hill plot, where $\hat{\alpha}_H$ is plotted against $k$, and look for a region where the graph has fairly stable behavior to identify the optimal value of the order statistics $k$. Alternatively, there are other methods for choosing the optimal $k$, such as the bootstrap approach \citep{Hall1990}.

In our simulations later in this paper we show that choosing $k$ from the Hill plot is very inaccurate and for higher values of $\alpha$ it is almost impossible to set its right value. \cite{Resnick2007} clearly illustrates why this technique very often leads to a ``Hill horror plot'' from which it is not possible to discern the correct value of $k$. Using the Hill estimates from simulated random variables on a standard symmetric L\'evy-stable distribution for different values of $k$, we also clearly demonstrate that choosing the ``correct'' $k$ is a very difficult task. Furthermore, for a short dataset ($<10^6$) it is usually impossible to find out the right value of $k$, especially when $\alpha$ is close to 2. Unfortunately, this setting is very often the case when we analyze empirical financial market data.

To overcome the problem of choosing the optimal $k$ we introduce a new estimation method based on comparison of the estimate of the empirical dataset with estimates on a pre-simulated random variable from the standard L\'evy distribution for $k$ on the interval [$1\% - 20\%$]. The main advantage of our approach is that it does not assume the smoothing parameter $k$ to be known. 

\section{Finite sample properties of the Hill estimator for different heavy tails}
We would like to study the finite sample behavior of the Hill estimator in detail for stable distributions with different heavy tails first. Let us start with an introduction to stable distributions.

\subsection{Stable distributions} 

Stable distributions are a class of probability laws with appealing theoretical properties. Their application to financial modeling comes from the fact that they generalize the Gaussian distribution, which does not describe well-known stylized facts about stock market data. Stable distributions allow for heavy tails and skewness. In this paper, we provide a basic idea about stable distributions. Interested readers can find the theorems and proofs in \cite{Nolan2003}, \cite{Zoltarev} and \cite{SamorodintskyTaquu1994}.
 
 The reason for the term stable is that stable distributions retain their shape up to scale and shift under addition: if $X$, $X_{1}$, $X_{2}$,..., $X_{n}$ are independent, identically distributed stable random variables, then for every $n$  
 \begin{equation}
\label{eq1}
X_{1}+X_{2}+...+X_{n}\overset{d}{=}c_{n}X+d_{n}
\end{equation}
for constants $c_{n}>0$ and $d_{n}$. Equality ($\overset{d}{=}$) here means that the right-hand and left-hand sides have the same distribution. Normal distributions satisfy this property: the sum of normals is normal. In general, the class of all laws satisfying (\ref{eq1}) can be described by four parameters, $(\alpha,\beta,\gamma,\delta)$. Parameter $\alpha$ is called the \textit{characteristic exponent} and must be in the range $\alpha\in(0,2]$. The coefficients $c_{n}$ are equal to $n^{1/\alpha}$. Parameter $\beta$ is called the \textit{skewness} of the law and must be in the range $-1 \leq \beta \leq 1$. If $\beta = 0$, the distribution is symmetric, if $\beta > 0$ it is skewed to the right, and if $\beta < 0$ it is skewed to the left. While parameters $\alpha$ and $\beta$ determine the
shape of the distribution, $\gamma$ and $\delta$ are scale and location parameters, respectively.

Due to a lack of closed form formulas for probability density functions (except for three stable distributions: Gaussian, Cauchy, and Levy) the $\alpha$-stable distribution can be described by a characteristic function which is the inverse Fourier transform of the probability density function, i.e., $\phi(u)=E\exp(iuX)$.

A confusing issue with stable parameters is that there are multiple parametrizations used in the literature. \cite{Nolan2003} provides a good guide to all the definitions. In this paper, we will use Nolan's parametrization, which is jointly continuous in all four parameters. A random variable $X$ is distributed by $S(\alpha,\beta,\gamma,\delta)$ if it has the following characteristic function:
 \begin{equation}
 \footnotesize
\label{eq2}
\phi(u)=
\left\{
\begin{array}{lr}
\exp(-\gamma^{\alpha}|u|^{\alpha}[1+i\beta(\tan\frac{\pi\alpha}{2})($sign$ u)(|\gamma u|^{1-\alpha}-1)]+i\delta u) & \alpha \ne 1 \\
\exp(-\gamma |u| [1+i\beta\frac{2}{\pi}($sign$ u) \ln (\gamma |u|)]+i\delta u) & \alpha = 1
\end{array}
\right.\end{equation}
There are only three cases where a closed-form expression for density exists and we can verify directly if the distribution is stable -- the Gaussian, Cauchy, and L\'evy distributions. Gaussian laws are stable with $\alpha=2$ and $\beta = 0$. More precisely, $N(0,\sigma^{2})=S(2,0,\sigma/\sqrt{2},0)$. Cauchy laws are stable with $\alpha=1$ and $\beta = 0$, Cauchy$(\gamma,\delta)=S(1,0,\gamma,\delta)$; and finally, L\'evy laws are stable with $\alpha=1/2$ and $\beta = 1$; L\'evy$(\gamma,\delta)=S(1/2,1,\gamma,\gamma+\delta)$. \cite{Nolan2003} shows these examples in detail.

For all values of parameter $\alpha<2$ and $-1<\beta<1$, stable distributions have two tails that are asymptotically power laws. The asymptotic tail behavior of non-Gaussian stable laws for $X\sim S(\alpha,\beta,\gamma,\delta)$ with $\alpha<2$ and $-1<\beta<1$ is defined as follows:
\begin{eqnarray}
\label{eq3}
\underset{x\rightarrow \infty }{\lim }x^{\alpha }P\left( X>x\right)=c_{\alpha }\left( 1+\beta \right) \gamma ^{\alpha } \\
\underset{x\rightarrow \infty }{\lim }x^{\alpha }P\left( X<-x\right)=c_{\alpha }\left( 1-\beta \right) \gamma ^{\alpha },
\end{eqnarray}
where
\begin{equation}
c_{\alpha }=\sin \left( \frac{\pi \alpha }{2}\right) \Gamma \left(\alpha \right) /\pi.
\end{equation}
If the data is stable, the empirical distribution function should be approximately a straight line with slope $-\alpha$ in a log-log plot.

A negative aspect of non-Gaussian stable distributions ($\alpha<2$) is that not all moments exist.\footnote{It is possible to define a fractional absolute moment of order $p$, where $p$ is any real number. For $0<p<\alpha$, $E\left\vert X\right\vert $ is finite, but for $p \ge \alpha$, $E\left\vert X\right\vert^{p}=\infty $ \citep{Nolan2003}.} The first moment $EX$ is not finite (or is undefined) when $\alpha \le 1$. On the other hand, when $1<\alpha \le 2$ the first moment is defined as
\begin{equation}
EX = \mu =\delta -\beta \gamma \tan \frac{\pi \alpha }{2}.
\end{equation}
Non-Gaussian stable distributions do not have finite second moment. It is also important to emphasize that the skewness parameter $\beta$ is different from the classical skewness parameter used for the Gaussian distribution. It cannot be defined because the second and third moments do not exist for non-Gaussian stable distributions. The kurtosis is also undefined, because the fourth moment does not exist either.

The characteristic exponent $\alpha$ gives important information about financial market behavior. When $\alpha < 2$, extreme events are more probable than for the Gaussian distribution. From an economic point of view some values of parameter $\alpha$ do not make sense. For example, in the interval $0 < \alpha < 1$ the random variable $X$ does not have a finite mean. In this case, an asset with returns which follow a stable law with $0 < \alpha < 1$ would have an infinite expected return. Thus, we are looking for $1 < \alpha < 2$ to be able to predict extreme values more precisely than by the Gaussian distribution.

\subsection{Research design}
The simulations are constructed so as to show how the Hill estimator behaves for different heavy tails data. For this purpose we use the L\'evy stable distribution depending on parameters $(\alpha,\beta,\gamma,\delta)$ and set them to $(\alpha,0,{\sqrt{2}}/{2},0)$, where $1.1 \le \alpha \le 2$ with a step of 0.1. For each parameter $\alpha$, 1,000 time series with lengths from $10^3$ to $10^6$ are simulated and the tail exponent is estimated using the Hill estimator for $k\le1\%$ and $k\in(1\%,20\%]$ separately.

In other words, we simulate the grid of different $\alpha$ for different series lengths. For each position in the grid, we estimate the tail exponent using the Hill estimator. This allows us to study the finite sample properties of the Hill estimator for different length series, different tail exponents and different $k$.

\subsection{Results from Monte Carlo simulations}
The biggest problem with the optimal choice of $k$ is that $\alpha$ is not known, so we cannot really choose the optimal $k$. In our simulation, we show the finite sample properties of the Hill estimator $k\le1\%$ and $k\in(1\%,20\%]$ separately.

\subsubsection{Hill estimation for $k>1\%$}
\begin{figure}[p]
   \centering
   \includegraphics[width=4.5in]{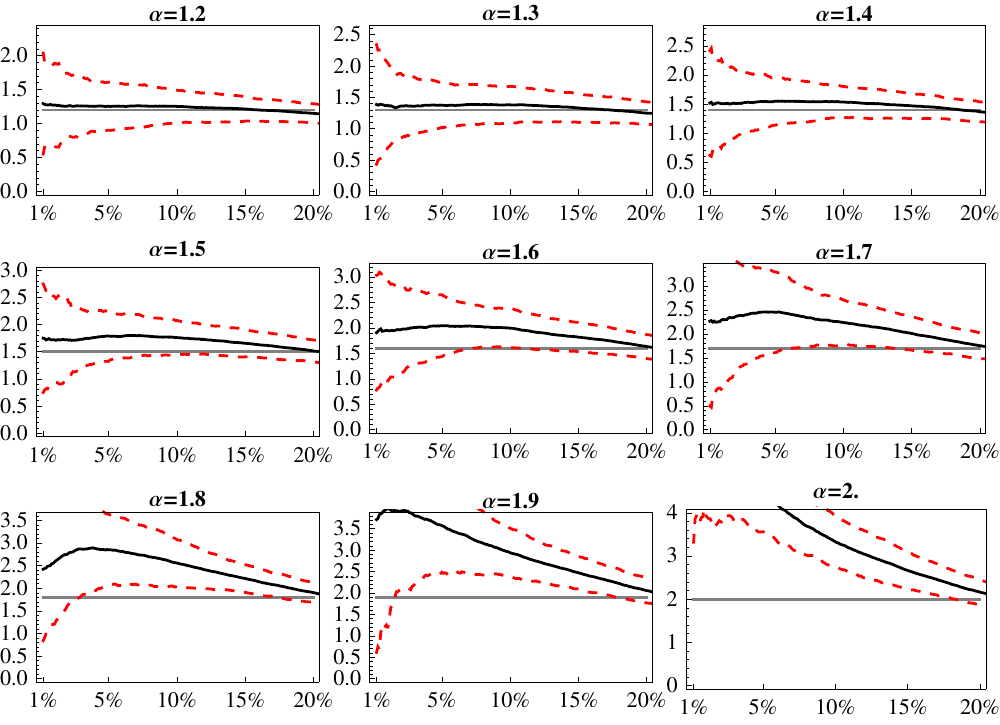} 
   \caption{Hill estimate of the tail exponent on the 1\%--20\% tail, $10^3$ length}
   \label{fig2}
\end{figure}

We begin the simulations with a time series length of $10^3$, which is a usual sample length for real-world financial data, i.e., it equates to approximately 4 years of daily returns of a stock. Figure \ref{fig2} shows the 95\% confidence intervals\footnote{All confidence intervals are computed from the asymptotic normality of the sample.} of the Hill estimate for different $\alpha$ for all $k$. It can be seen that it is very difficult to statistically distinguish between different tail exponents $\alpha$ and it is very unclear which optimal $k$ we should choose. 

\begin{figure}[p]
   \centering
   \includegraphics[width=4.5in]{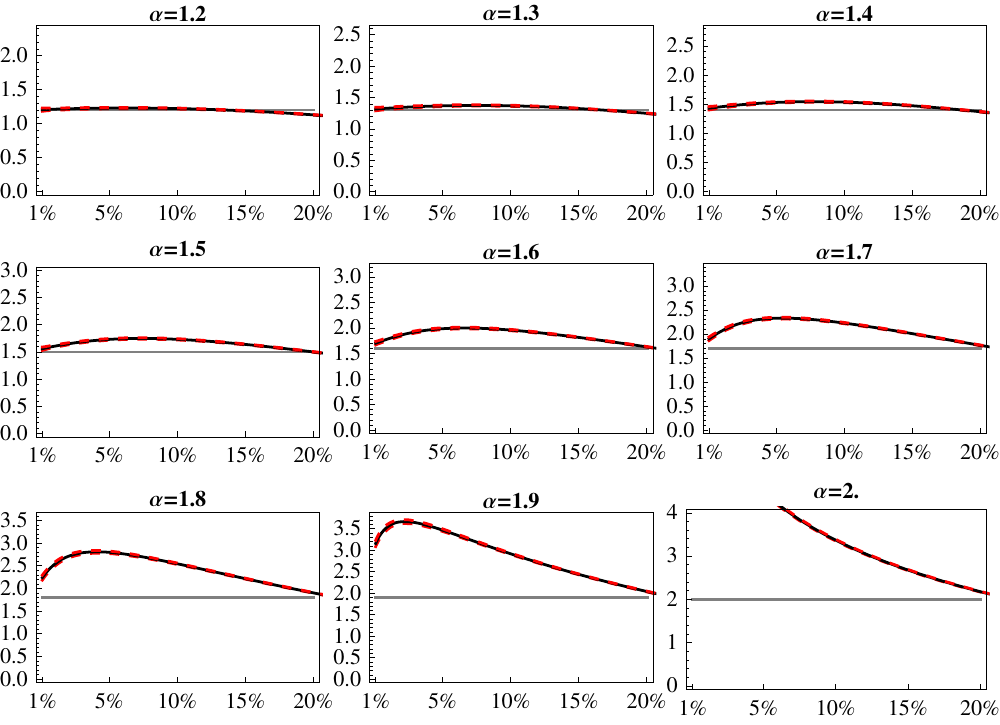} 
   \caption{Hill estimate of the tail exponent on the 1\%--20\% tail, $10^6$ length}
   \label{fig4}
\end{figure}

Figure \ref{fig4} shows much more precise results for a time series of length $10^6$. This allows for narrow confidence intervals, but even with this exact result we can see that the Hill estimator is problematic as it overestimates the value of the tail exponent and we cannot pick the optimal $k$ even from this large grid of simulated data. The only way of estimating the tail exponent would be to pick $k$, estimate the tail exponent, and compare it to the grid of simulated confidence intervals. But we would still need at least a $10^6$ sample size to achieve an exact result. As we can see in Figure \ref{fig2}, we cannot get a statistically significant estimate on data with length $10^3$. Table \ref{taboptimalk20} shows the optimal value of $k$, where the Hill estimator gives a good estimate of the tail exponent $\alpha$\footnote{Optimal value of $k$ is such $k$ for which the difference of the estimated and simulated $\alpha$ is minimal.}.
 
\begin{table}[h]
\footnotesize
\begin{center}
\begin{tabular}{@{} ccccccccccc @{}}
\toprule
& \multicolumn{10}{c}{$\alpha$} \\
\cmidrule{2-11}
 n & 1.1 & 1.2 &1.3 & 1.4 & 1.5 & 1.6 & 1.7 & 1.8 & 1.9 & 2 \\
\midrule
 $10^3$ & 9.11\% & 15.92\% & 17.28\% & 18.53\% & 20.42\% & 20.84\% & 21.15\% & 21.78\% & 22.20\% & 21.88\% \\
 $10^4$ & 8.88\% & 13.55\% & 17.17\% & 18.73\% & 19.95\% & 20.83\% & 21.38\% & 21.71\% & 21.97\% & 22.01\% \\
 $10^5$ & 8.04\% & 13.73\% & 16.84\% & 18.78\% & 20.04\% & 20.80\% & 21.35\% & 21.73\% & 21.98\% & 22.10\% \\
 $10^6$ & 8.29\% & 13.60\% & 16.80\% & 18.74\% & 20.00\% & 20.80\% & 21.35\% & 21.73\% & 21.98\% & 22.15\% \\
\bottomrule
\end{tabular}
\caption{Optimal $k$ for various sample lengths $n$ from $10^3$ to $10^6$}
\label{taboptimalk20}
\end{center}
\end{table}

The figures suggest that the Hill estimator does not overestimate the tail exponent on the 1\% tail so strongly. Thus, we repeat the exercise for $k\le1\%$.

\subsubsection{Hill estimation for $k\le1\%$}
\begin{figure}[h!]
   \centering
   \includegraphics[width=4.5in]{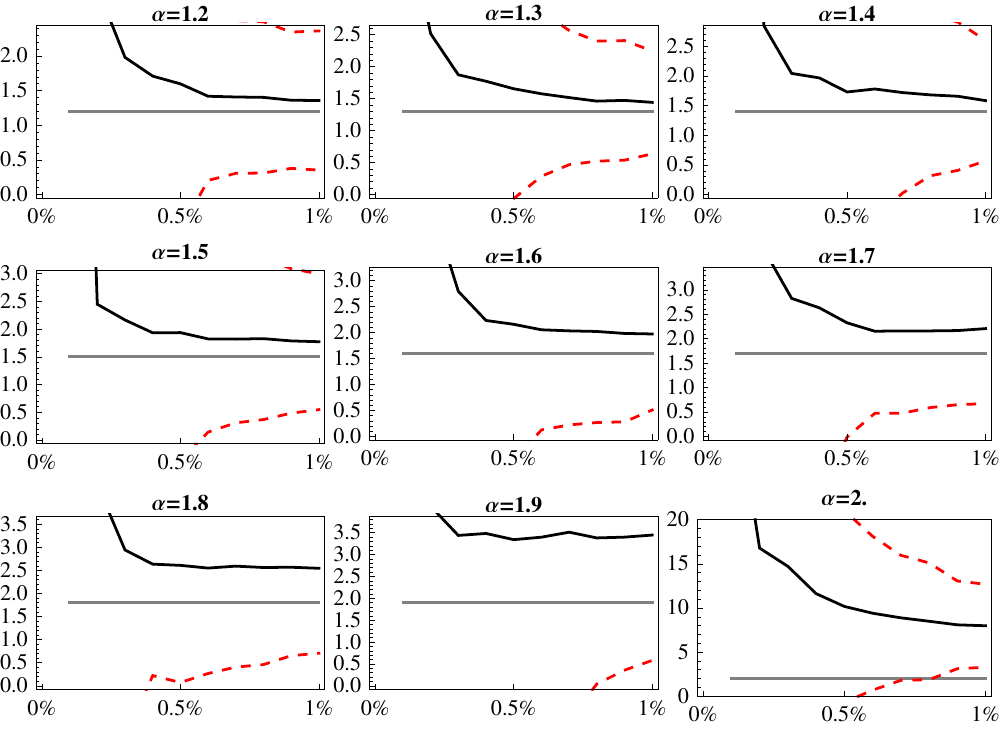} 
   \caption{Hill estimate of the tail exponent on the $\le1\%$ tail, $10^3$ length}
   \label{fig5}
 
   \includegraphics[width=4.5in]{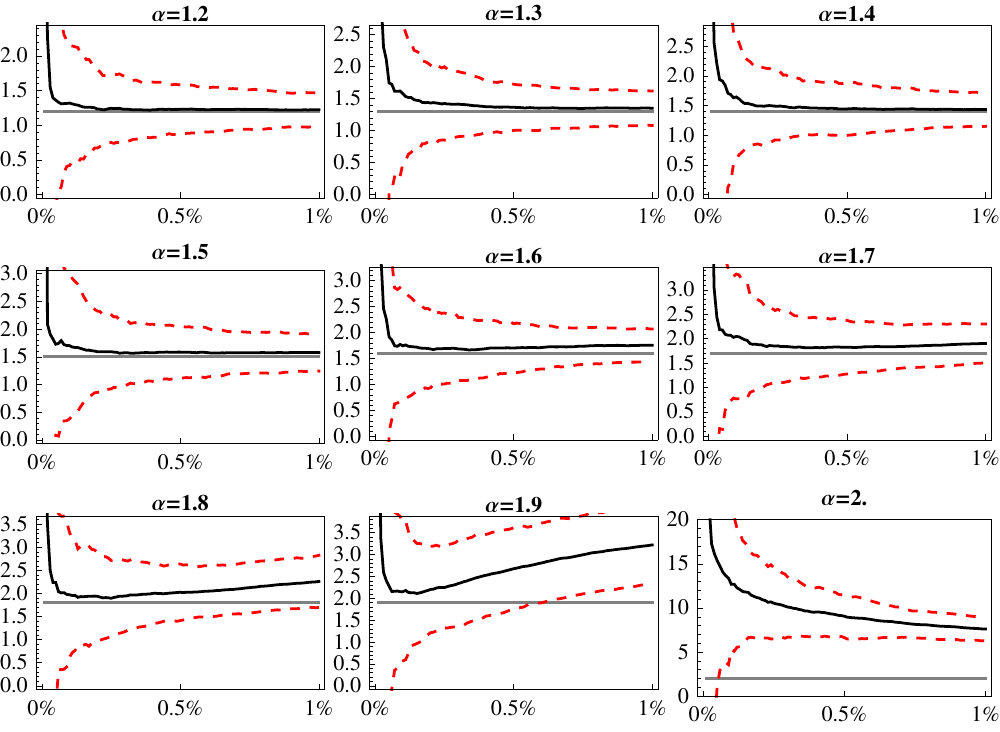} 
   \caption{Hill estimate of the tail exponent on the $\le1\%$ tail, $10^4$ length}
   \label{fig6}
\end{figure}

\begin{figure}[h!]
   \centering
   \includegraphics[width=4.5in]{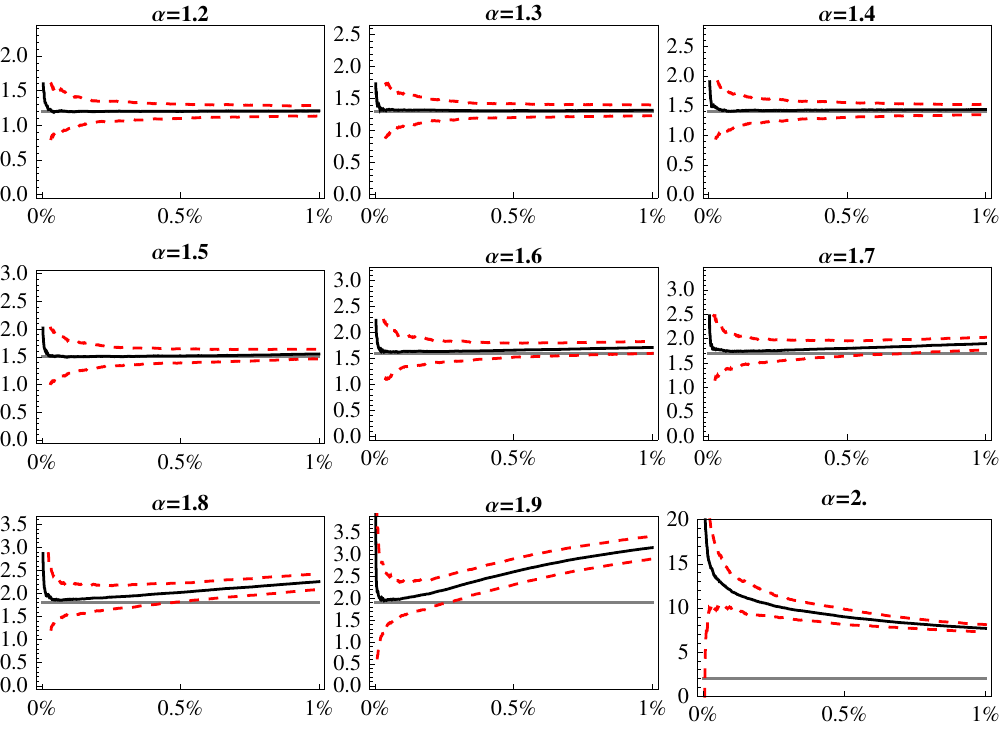} 
   \caption{Hill estimate of the tail exponent on the $\le1\%$ tail, $10^5$ length}
   \label{fig7}
   \centering
   \includegraphics[width=4.5in]{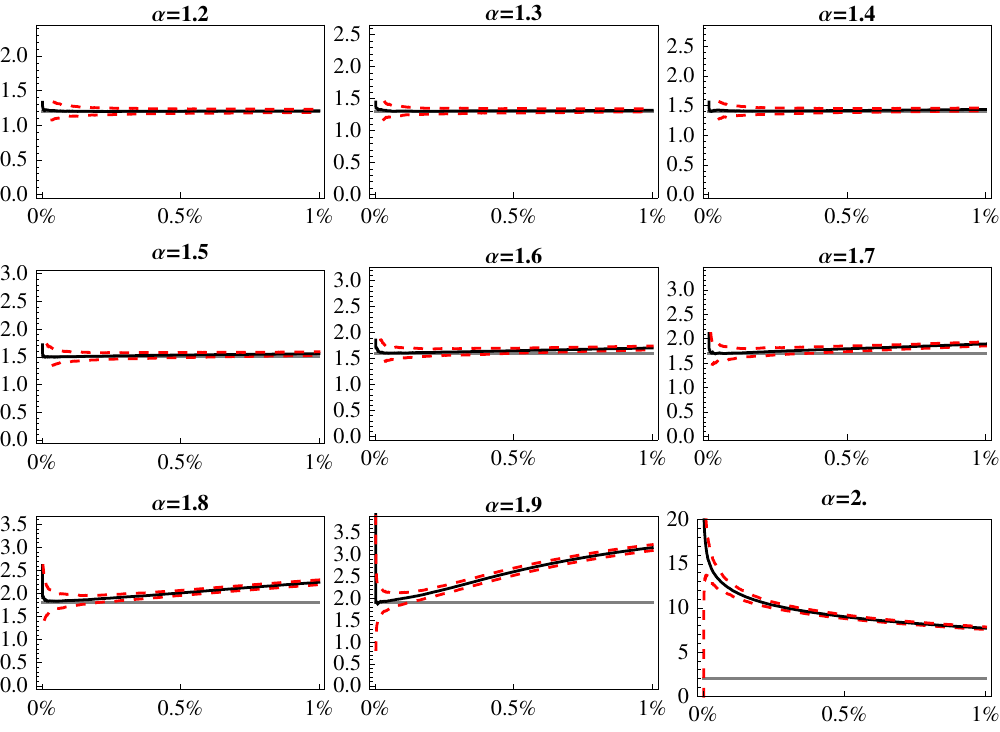} 
   \caption{Hill estimate of the tail exponent on the $\le1\%$ tail, $10^6$ length}
   \label{fig8}
\end{figure}

Figures \ref{fig5}, \ref{fig6}, \ref{fig7}, and \ref{fig8} show the behavior of the Hill estimator on generated datasets with length $10^3$ up to $10^6$. The results suggest that it makes no sense to use the Hill estimator for tail exponent estimation even on the $10^5$ dataset. For example, $\alpha$ of $1.2, 1.3, 1.4, 1.5$, and $1.6$ cannot be statistically distinguished from each other. Only the dataset with length $10^6$ shows better results. Using the simulations on such a long dataset, we can choose the optimal range of $k$. More exactly, the optimal $k$ interval is such where the estimated value of the tail exponent does not deviate from its theoretical value by more than 5\%. Table \ref{tab1} shows the optimal sets of $k$ where the Hill estimator was a good estimate of the tail exponent. 
 
 \begin{table}[h]
\footnotesize
\begin{center}
\begin{tabular}{@{} ccc @{}}
\toprule
$\alpha$ & & Range \\
\midrule
1.2 & & $0.08\% - 1\%$ \\
1.3 & &  $0.1\% - 0.35\%$ \\
1.4 & &  $0.06\% - 0.22\%$ \\
1.5 & &  $0.15\% - 0.25\%$ \\
1.6 & &  $0.01\% - 0.12\%$ \\
1.7 & &  $0.03\% - 0.12\%$ \\
1.8 & &  $0.02\% - 0.08\%$ \\
1.9 & &  $0.02\% - 0.06\%$ \\
2 & &  $0.005\% - 0.03\%$ \\
\bottomrule
\end{tabular}
\caption{Optimal range of $k$ for series length $10^6$ for $k\le1\%$ }
\label{tab1}
\end{center}
\end{table}
 
In order to estimate the true tail exponent, we need a dataset with length of at least $10^6$, otherwise inference of the tail exponent may be strongly misleading and rejection of the L\'evy stable regime not appropriate. For the sake of clarity we have to note that the results for the $\alpha=2$ should be interpreted with caution as it is the special case of Gaussian distribution which does not have heavy tails. Thus the Hill estimator is not appropriate in this case.
 
In the next section, we develop a Monte Carlo-based tail exponent estimator which deals with all of these problems.

\section{Tail exponent estimator based on Monte Carlo simulations}
After the study of the finite sample properties of the Hill estimator, we utilize the results and introduce our Monte Carlo-based estimator. The Monte Carlo technique provides an attractive method of building exact tests from statistics whose finite sample distribution is intractable but can be simulated, thus it can be utilized in our problem. 

We construct our Monte Carlo-based tail exponent estimator as follows.
\begin{enumerate}
\item Generate 1,000 $i.i.d.$ $\alpha-$stable distributed random variables $X^{\alpha_0}$ of length $n$, i.e. $x_1^{\alpha_0},x_2^{\alpha_0},\ldots,x_n^{\alpha_0}\sim S(\alpha_0,0,\sqrt{2}/2,0)$, for each $\alpha_0$ parameter from the range $\left[1.01,2\right]$ with step $0.01$.
\item For each $X^{\alpha_0}$, estimate the tail exponent using the Hill estimator for all $k$ from the interval $(1\%-20\%)$, i.e. $\hat{\alpha}_{\alpha_0,k}$.
\item From Monte Carlo simulations compute the expected value $E[\hat{\alpha}_{\alpha_0,k}]$ of the Hill estimator for all $k$ and all $\alpha_0$.
\item Using the Hill estimator estimate the tail exponent $\hat{\alpha}_{emp,k}$ on an empirical dataset of length $n$ for all $k$ from the interval $(1\%-20\%)$. 
\item Our Monte Carlo-based estimator $\hat{\alpha}_{MC}$ is defined as: \\
\begin{equation}
\hat{\alpha}_{MC}=\underset{\alpha_0 \in[1.01,2]}{\operatorname{\arg\,min}} \sum_{k}|\hat{\alpha}_{emp,k}-E[\hat{\alpha}_{\alpha_0,k}]|
\label{eq:MC}
\end{equation}
\end{enumerate}

In other words, we simulate random variables $X$ of length $n$ for 100 different $\alpha$'s (representing the tail parameter) 1,000 times.\footnote{A more exact confidence interval can be obtained by generating more random variables} On these random variables we estimate the tail exponent using the Hill estimator on the $(1\%-20\%)$ tail (as we have shown in the previous section, the Hill estimator does not really work on $k\le1\%$ for series of lengths $\le10^6$). Using the simulated variables we get the expected value of the Hill estimator for all $k$ and all $\alpha_0$.

After we have obtained the behavior of the Hill estimator for a large grid of different tail exponents and $k$, we can estimate the tail exponent on the empirical dataset. Thus, we estimate the tail exponent of empirical data using the Hill estimator for all values of $k$. As the last step, we minimize the loss function which is defined by Equation \ref{eq:MC} through the whole simulated grid of $\alpha_0$. Our Monte Carlo-based tail exponent estimate $\hat{\alpha}_{MC}$ is the value for which the distance between $E[\hat{\alpha}_{\alpha_0,k}]$ and $\hat{\alpha}_{emp,k}$ is minimal (Equation \ref{eq:MC}).

\subsection{Properties of $\hat{\alpha}_{MC}$}

In this section we study the finite sample properties of our estimation method. For this purpose, we will 
again use the Monte Carlo procedure, which will provide us the statistical inference for our estimator. We simulate independent identically $\alpha-$stable distributed random variables of length $n$ from $S(\alpha,0,\sqrt{2}/2,0)$ with varying $\alpha \in[1.01,2]$ with step $0.01$. For each simulated sample, we apply our method and estimate the tail exponent. This Monte Carlo simulation allows us to derive the finite sample properties of our estimator, so we can get the confidence interval of the estimates. 

\begin{figure}
   \centering
   \includegraphics[width=2.5in]{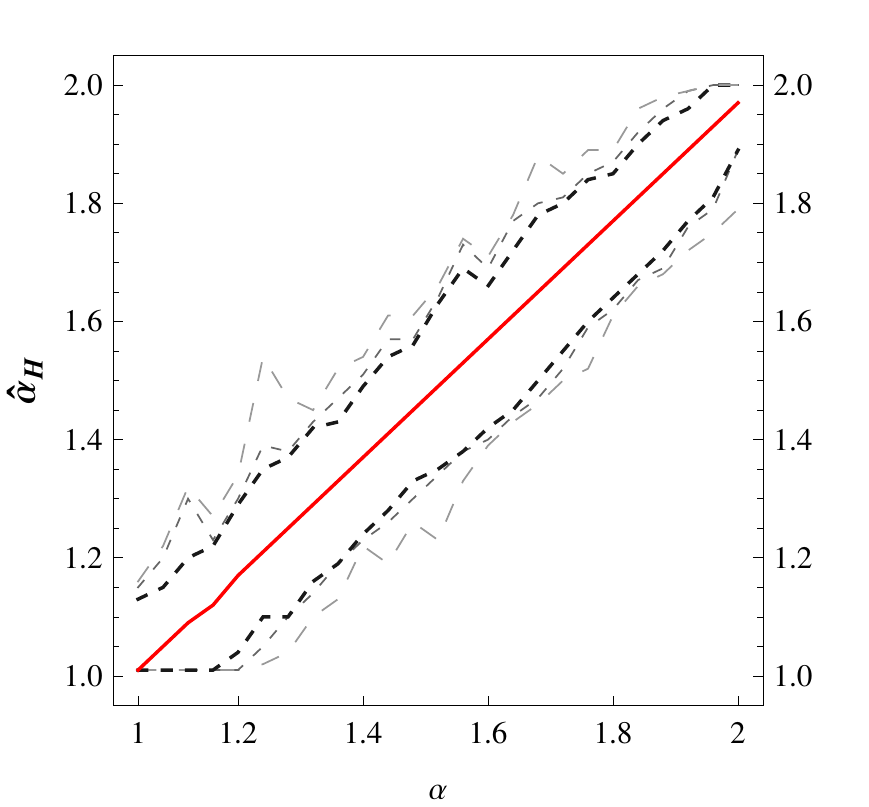} 
   \caption{Plots of  99\%, 95\%, and 90\% quantiles of our Monte Carlo-based tail exponent estimator based on 100 simulations.}
   \label{figconf}
\end{figure}

Figure \ref{figconf} plots the 99\%, 95\%, and 90\% quantiles of our tail exponent estimator based on 100 simulations. Table \ref{tabCI} provides exact simulated confidence intervals for several values of $\alpha$.

\begin{table}
\footnotesize
\begin{center}
\begin{tabular}{@{} lccccccccc @{}}
\toprule
& & $0.5\%$ & $2.5\%$ & $5\%$ &$\hat{\alpha}_{MC}$ & $95\%$ & $97.5\%$ & $99.5\%$ \\
\toprule
 $\alpha=1.1$ & & 1.01 & 1.01 & 1.01 & 1.11 & 1.21 & 1.22 & 1.25 \\
 $\alpha=1.2$ & & 1.01 & 1.04 & 1.05 & 1.19 & 1.31 & 1.34 & 1.38 \\
 $\alpha=1.3$ & & 1.12 & 1.13 & 1.15 & 1.29 & 1.41 & 1.42 & 1.49 \\
 $\alpha=1.4$ & & 1.23 & 1.26 & 1.27 & 1.39 & 1.54 & 1.56 & 1.58 \\
 $\alpha=1.5$ & & 1.34 & 1.35 & 1.37 & 1.49 & 1.61 & 1.63 & 1.66 \\
 $\alpha=1.6$ & & 1.38 & 1.48 & 1.50 & 1.60 & 1.77 & 1.80 & 1.88 \\
 $\alpha=1.7$ & & 1.55 & 1.56 & 1.58 & 1.69 & 1.79 & 1.83 & 1.87 \\
 $\alpha=1.8$ & & 1.62 & 1.62 & 1.66 & 1.79 & 1.88 & 1.89 & 1.96 \\
 $\alpha=1.9$ & & 1.77 & 1.77 & 1.78 & 1.89 & 1.98 & 2 & 2 \\
 $\alpha=2$ & & 1.94 & 1.95 & 1.95 & 2 & 1.99 & 2 & 2 \\
\bottomrule
\end{tabular}
\caption{Quantiles of our estimation method.}
\label{tabCI}
\end{center}
\end{table}

In comparison with the Hill estimator, or any other tail exponent estimator, our method has several advantages. The largest advantage is that the method is not sensitive to the choice of $k$, unlike most of the other estimation methods discussed in the previous text. Our method also works well for smaller samples, i.e. $10^3$, as it yields much narrower confidence intervals compared to the Hill estimator.

 \section{Empirical study}
To illustrate the power of our tail exponent estimation method, we employ empirical data. We use the daily returns of the following stock market indices: the German DAX 30, the U.S. S\&P 500 and Dow Jones Industrial 30 (DJI), the London FTSE 100, the Nikkei 225, and the Singapore Straits Times from the beginning of 2002 to the end of 2009. More precisely, we choose the last 2,000 observations for each index and divide them into two equal periods, 2002--2005 and 2006--2009, each containing 1,000 observations. This way, we can compare different world stock market indices and their behavior before and during the current financial crisis .

Figures \ref{fig01} and \ref{fig02} show the histograms of the tested indices. Figure \ref{fig01} shows the histograms for the first period 2002--2005, while Figure \ref{fig02} shows the histograms for the second period 2006--2009. All the data are leptokurtic, showing excessive peaks around the mean and thicker tails than those of the normal density. Moreover, the data from the second period tend to show even more excessive kurtosis and heavier tails than the ones from the first period. This is probably caused by the large price movements that occurred during the deep financial turmoil in the years 2007--2009. By contrast, the first period of stable growth is closer to the standard normal distribution.

We use our Monte Carlo method to estimate the tail exponents for all the datasets. Before we start the estimations, we set the length $n$ of the simulated series from our algorithm to 1,000 so that it corresponds to the length of the empirical dataset and we can statistically compare the estimates. Table \ref{tab2} shows the tail exponent estimates with confidence intervals.
\begin{table}[h]
\footnotesize
\begin{center}
\begin{tabular}{@{} lccccccccc @{}}
\toprule
& &\multicolumn{7}{c}{Empirical Confidence Intervals} \\
\cmidrule{3-9}
& & $0.5\%$ & $2.5\%$ & $5\%$ &$\hat{\alpha}_{MC}$ & $95\%$ & $97.5\%$ & $99.5\%$ \\
\toprule
 \text{German DAX 30}  & 2002--2005	 & 1.50 & 1.53 & 1.55 & \textbf{1.64} & 1.77 & 1.8 & 1.85 \\
 		 				& 2006--2009 	& 1.47 & 1.53 & 1.56 & \textbf{1.70 }& 1.79 & 1.81 & 1.87 \\
\midrule
 \text{U.S. S\&P500} 	& 2002--2005	 & 1.53 & 1.57 & 1.58 & \textbf{1.74} & 1.85 & 1.88 & 1.94 \\
 		 				& 2006--2009 	& 1.35 & 1.38 & 1.4 &\textbf{1.54} & 1.63 & 1.70 & 1.72 \\
\midrule
 \text{U.S. DJI 30} 		& 2002--2005	 & 1.50 & 1.53 & 1.55 & \textbf{1.69} & 1.83 & 1.87 & 1.89 \\
 		 				& 2006--2009 	& 1.29 & 1.38 & 1.4 & \textbf{1.54} & 1.66 & 1.70 & 1.77 \\
\midrule
 \text{London FTSE 100} & 2002--2005	 & 1.34 & 1.42 & 1.43 & \textbf{1.57} & 1.69 & 1.69 & 1.71 \\
 		 				& 2006--2009 	&1.50 & 1.55 & 1.55 & \textbf{1.69} & 1.82 & 1.87 & 1.91 \\
\midrule
 \text{Japan Nikkei 225}  & 2002--2005	& 1.79 & 1.81 & 1.84 & \textbf{1.93} & 1.98 & 2 & 2 \\
 		 				& 2006--2009 	 & 1.58 & 1.60 & 1.62 & \textbf{1.77} & 1.88 & 1.89 & 1.89 \\
\midrule
 \text{Singapore Straits} & 2002--2005	 & 1.72 & 1.72 & 1.74 & \textbf{1.85} & 1.94 & 1.95 & 1.97 \\
  		 					& 2006--2009 	& 1.47 & 1.50 & 1.52 & \textbf{1.66} & 1.78 & 1.81 & 1.87 \\
\bottomrule
\end{tabular}
\caption{Exact confidence intervals (quantiles) of the estimated tail exponent $\alpha$.}
\label{tab2}
\end{center}
\end{table}
\begin{figure}[h!]
   \centering
   \includegraphics[width=5in]{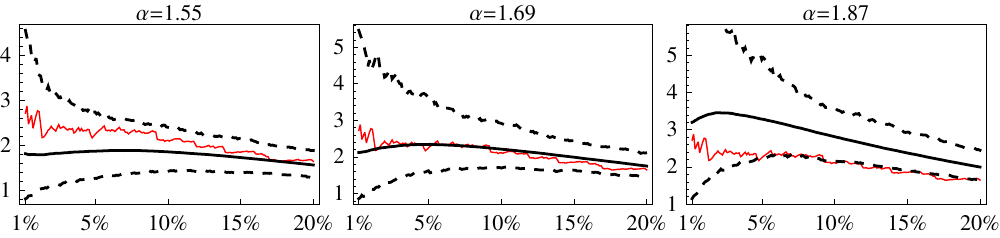} 
   \caption{Plot of Hill estimates through different $k$ for the FTSE index (second period of 2006--2009) for $\alpha=1.55, \alpha=1.69$, and $\alpha=1.87$, which are our $\hat{\alpha}_{MC}$ with the 0.025\% and 97.5\% quantiles.}
   \label{fig1}
\end{figure}
Let us further demonstrate the estimation method on the example of the London FTSE 100 index for the second period. Figure \ref{fig1} plots the empirical Hill estimates through different $k$ and compares them to 1,000 simulations of $\alpha=1.55$, $\alpha=1.69$, and $\alpha=1.87$ (these are our Monte Carlo estimates with the 0.025\% and 97.5\% quantiles). It is immediately visible from the plots that the loss function defined in our estimation algorithm reaches its minimum for $\alpha=1.69$. We plot another two cases just for illustration of what our estimate looks like at the critical levels. After minimizing the loss function and choosing the tail exponent, we derive the finite sample properties of the estimate. From Table \ref{tab2}, we can see that 1.55 and 1.87 are the 0.025\% and 97.5\% quantiles.

Unlike the classical Hill estimation procedure, we can see that our method is not sensitive to the choice of $k$. Moreover, it is a much stronger result for the tail exponent, as it accounts for all possible $k$'s, not only the one chosen. Using the classical Hill estimation procedure, we would select $k$ and compute the tail exponent, but as we discussed in the previous text, there is always a trade-off between bias and variance when choosing the optimal $k$. Our method reduces this problem. Finally, the sample on which we estimate the tail exponent is quite small. In previous sections we showed that Hill estimation does not work well for such small samples, but our method works fine.
  
\section{Conclusion}
In this paper, we have researched the finite sample behavior of the Hill estimator under $\alpha-$stable distributions and we introduce a Monte Carlo-based method of estimation for the tail exponent.
 
First of all, we provide Monte Carlo confidence intervals for the Hill estimator under different $\alpha$'s and $k$ on the interval $k\in(0,20\%]$ of sample size $10^3$ up to $10^6$. We also provide the optimal range of $k$ for the series length $10^6$. Based on our simulations, we conclude that in order to estimate the true tail exponent using the Hill estimator, we need a dataset of at least $10^6$ length, i.e., high frequency data, otherwise inference of the tail exponent may be strongly misleading and rejection of the L\'evy stable regime not appropriate.
 
In the second part of the paper, we utilize the results and introduce a tail exponent procedure based on Monte Carlo simulations. It is based on the idea of simulating a large grid of random variables from the L\'evy stable distribution with different $\alpha$ exponents and estimating the tail exponent using the Hill estimator for all different $k$. After obtaining the behavior of the Hill estimator on this large grid of different tail exponents and $k$, all we need to do is to estimate the tail exponent on the empirical dataset for all $k$'s and compare all the values with the simulated grid. Using this algorithm we get estimates of the tail exponent.
 
In comparison with the Hill estimator, or any other tail exponent estimator, our method has several advantages. The largest advantage is that it is not sensitive to the choice of $k$. Moreover, the method works well for small samples. The new estimator gives unbiased results with symmetrical confidence intervals.
  
Finally, we illustrate the power of our tail exponent estimation method on an empirical dataset. We use daily returns of leading world stock markets: the German DAX, the U.S. S\&P 500 and Dow Jones Industrial (DJI), the London FTSE, the Nikkei 225, and the Singapore Straits Times from the beginning of 2002 to the end of 2009. We divide this period into two equal sub-periods and compare the tail exponents estimated using our method. \\

\bibliography{reference_PhyA_Hill}
\bibliographystyle{chicago}

\newpage
\section*{Appendix}

\begin{figure}[h]
   \centering
   \includegraphics[width=4.5in]{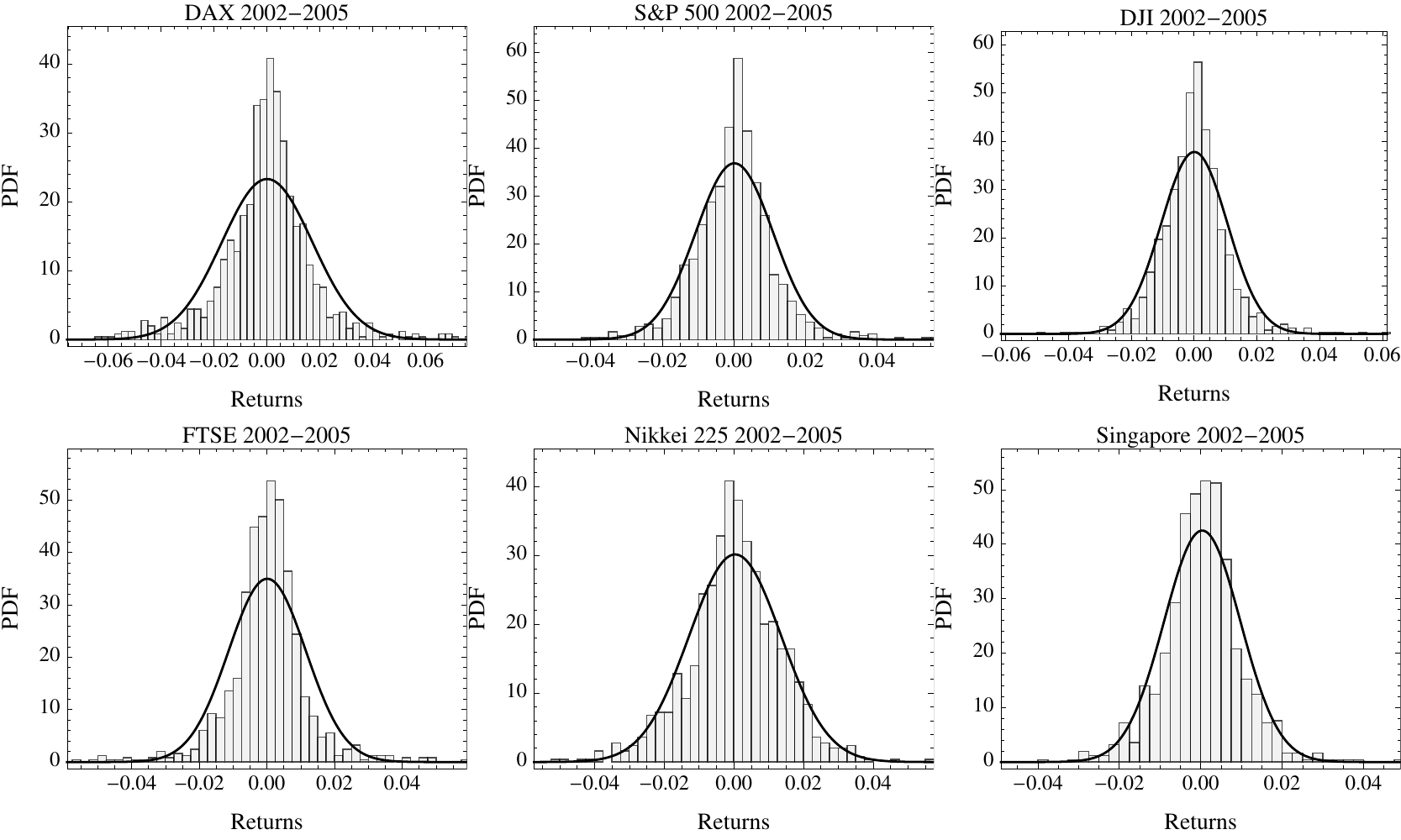} 
   \caption{Histograms of all indices for the first period compared with the standard normal distribution}
   \label{fig01}
\end{figure}

\begin{figure}[h]
   \centering
   \includegraphics[width=4.5in]{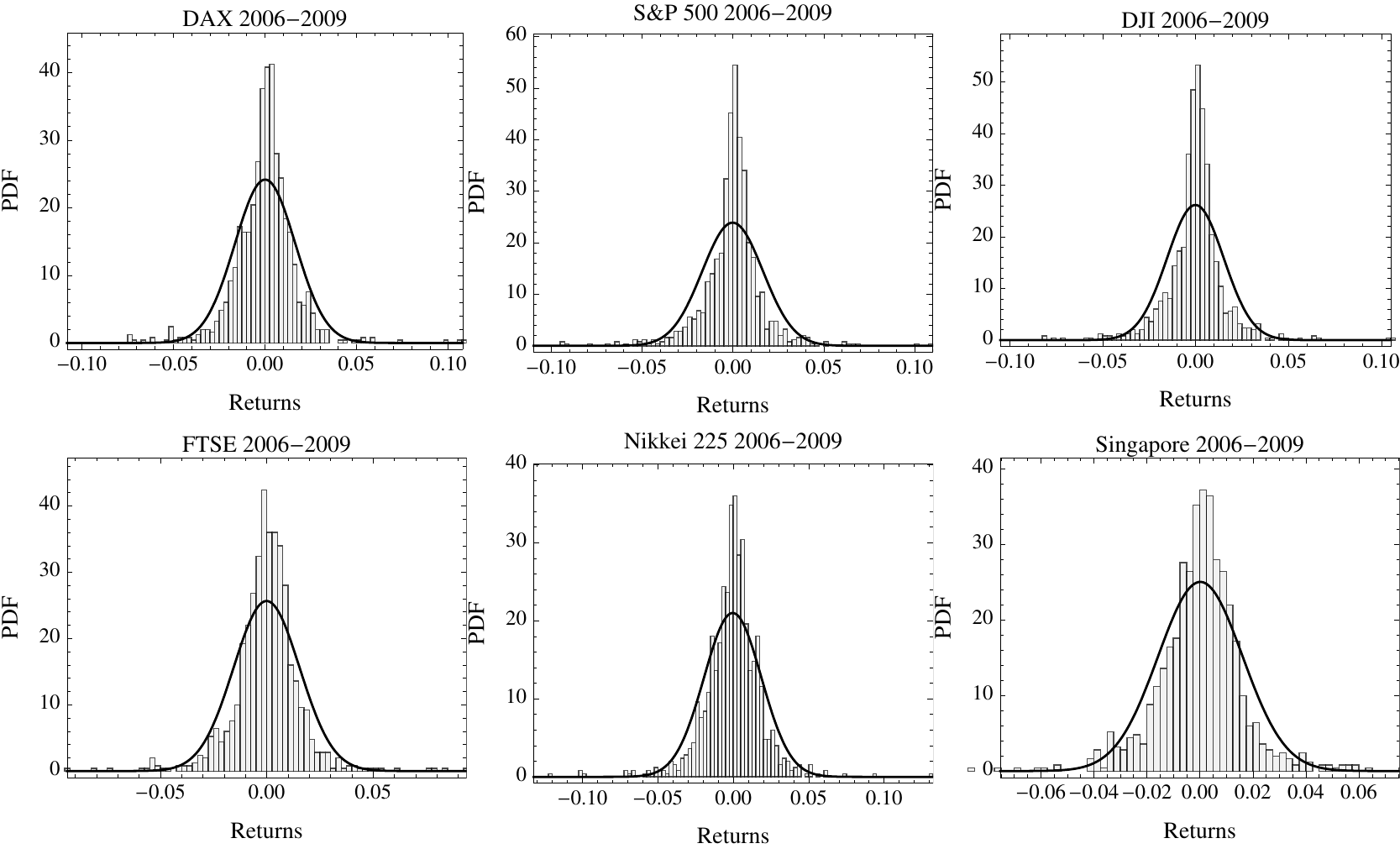} 
   \caption{Histograms of all indices for the second period compared with the standard normal distribution}
   \label{fig02}
\end{figure}

\end{document}